\documentstyle[prb,twocolumn,aps]{revtex}

\begin{document}
\tolerance 50000

\draft

\title{Evidence for a superfluid density in  $t$--$J$ ladders}
\author{C.A.Hayward$^{1}$, D.Poilblanc$^{1}$, R,M.Noack$^{2}$,
 D.J.Scalapino$^{3}$, W.Hanke$^{2}$
}
\address{
$^{1}$Lab. de Physique Quantique, Universit\'e Paul Sabatier,
31062 Toulouse, France \\
$^{2}$Institute for Physics, University of Wurzburg, D-8700 Wurzburg, Germany\\
$^{3}$Department of Physics, University of California Santa Barbara, CA 93106
\\
}

\twocolumn[
\date{March 95}
\maketitle
\widetext

\vspace*{-1.0truecm}

\begin{abstract}
\begin{center}
\parbox{14cm}{
Applying three independent techniques, we give numerical evidence for a finite
superfluid density in isotropic hole-doped
t--J ladders: We show the existence of anomalous flux quantization, emphasising
the
contrasting behaviour to that found in the `Luttinger liquid' regime stabilised
at
low electron densities; We  consider the nature
of the low-lying  excitation modes, finding the 1-D analog of the
superconducting state; And using a density matrix renormalization group
approach, we find long range pairing correlations and exponentially decaying
spin-spin correlations.
}
\end{center}
\end{abstract}

\pacs{
\hspace{1.9cm}
}
]
\narrowtext

The behaviour of strongly correlated electrons confined to
coupled chains is at present a topic undergoing much investigation; the
reasons for this attention are numerous.
Firstly, with the behaviour of electrons under t--J or Hubbard type
interactions in one-dimension now  relatively well
understood, the two-chain systems provide an interesting `first step' towards
the challenge of
two-dimensions.
Secondly, the unusual nature of the ground state  of the  undoped system,
in particular the existence of a spin gap \cite{heisladder},
leads to further interest with regards to   `gapped' superconducting
behaviour.
Furthermore, it is believed that  compounds such as (VO$_2$)P$_2$O$_7$
\cite{compounds1}{
and SrCu$_2$O$_3$ \cite{compounds2} may be described
by a lattice of  coupled chains.
Whilst there is considerable literature concerning
the possible  phases in a  t--J ladder, a complete picture is still far from
being realised and our aim is to clarify
the behaviour in a particular region, specifically considering
the nature of the gapped state
when the system is doped.
 Various techniques have been
applied previously  in this hole doped region \cite{scmethods}
and there is some indication
for hole pairing and modified d-wave superconducting correlations.

The $t$-$J$ Hamiltonian on the $2\times L$ ladder is defined as,
\begin{eqnarray}
   {\cal H}=
   J^\prime \; \sum_{j}
   ({\bf S}_{j,1} \cdot {\bf S}_{j,2}
   - \textstyle{1\over4} n_{j,1} n_{j,2} ) \cr
   +J\;\sum_{\beta,j}
   ({\bf S}_{j,\beta} \cdot {\bf S}_{j+1,\beta}
   - \textstyle{1\over4} n_{j,\beta} n_{j+1,\beta} ) \cr
      - {t} \sum_{j,\beta ,s}
   P_G({c}^{\dagger}_{j,\beta;s}{c}_{j+1,\beta;s} + {\rm H.c.})P_G  \cr
      - {t^\prime} \sum_{j,s}
   P_G({c}^{\dagger}_{j,1;s}{c}_{j,2;s} + {\rm H.c.})P_G ,
\label{hamiltonian}
\end{eqnarray}
\noindent
where most notations are standard. $\beta$ (=1,2)
labels the two legs of the ladder (oriented along the
$x$-axis) while $j$ is a rung index ($j$=1,...,$L$).
We shall concentrate on the isotropic case where
the intra-ladder
 (along $x$) couplings   J and t  are equal to
the inter-ladder   (along $y$) couplings   $J^\prime$ and $t^\prime$.

At half filling, the hamiltonian reduces to the Heisenberg model and  the
behaviour
is generally
relatively well
understood \cite{heisladder}.
A  simple interpretation  is given by considering the
strong coupling limit ($J=0$);
in such a limit, the ground state consists of
a  singlet  on each rung with a spin gap  ($\sim J^\prime$) which
corresponds to forming a triplet on one of the rungs. With the introduction of
interchain coupling $J$, the triplets  can propagate and form a coherent band
thereby reducing the spin gap. In the isotropic  case
 the gap remains ($\sim 0.5J$) and it is the nature of the state formed
 on doping such a system that we shall concentrate on.

A possible phase diagram for the isotropic
t--J ladder as a function of $J/t$ and doping
has been  proposed recently \cite{DP}, the
main features of which we summarize.
Away from half filling,
the spin gapped region persists, exhibiting hole pairing and, as we will show,
  possible
superconducting correlations.
This behaviour is observed for a large region of $J/t$ except for
  $J/t$ $>\sim 2.5$ \cite{tsunetsugu}
  where  the system phase separates and also (perhaps) very small
 $J/t$ where the gap may disappear.
As the system is doped further, a
Luttinger-like phase is stabilised exhibiting gapless spin and charge
excitations. At very small electron densities, an electron paired phase exists.

In this letter  we will describe three {\it independent}
forms of evidence for a finite superfluid density in the spin gapped region
of the phase diagram (we will work specifically with an
electron density $\big <n \big > =0.8$).
The first set of results, and the ones we deal
in most detail with, are based on the existence  of
anomalous flux quantization, a feature present  in a superconducting state.
Secondly,
we consider  the spin and charge excitation modes which may be used to
characterize the
state of a system. Finally, we present direct calculations of correlation
functions
which have been obtained using the density matrix renormalization group method.

Our first set of results then concern
the existence  of anomalous flux quantization.
The calculation involves threading the double chain ring with a flux $\Phi$
and studying the functional form of the   ground
state energy with
respect to the threaded flux, namely  $E_0$($\Phi$); throughout this letter
we will measure the flux $\Phi$
in units of the flux quantum $\Phi_0$=$hc/e$. In general, $E_0$($\Phi$)
consists
of an envelope of a series of parabola, corresponding to the curves of
individual many body states $E_n$($\Phi$), exhibiting a  periodicity of one.
Byers and Yang \cite{byers_yang} have shown that in the thermodynamic limit,
$E_0$($\Phi$)
exhibits local minima at quantized values of  $\Phi$, the separation of which
is
$1$/$n$ where  $n$ is the
sum of charges in the basic group;
these local minima in $E_0$($\Phi$)
must be separated by a finite energy barrier.
 Hence, for
a superconductor we would expect minima in $E_0$($\Phi$) at intervals of
$1/2$;
these minima are related to the existence of
supercurrents which are  trapped in the metastable states corresponding to the
flux minima
and are thus unable to  decay away \cite{Schrieffer}. This phenomenon is
known as anomalous flux quantization.
It should be mentioned that the existence of anomalous flux quantization
is an indication of
pairing and is not sufficient in itself to imply a superconducting state.

Detailed studies of the attractive
Hubbard model  on two-dimensional lattices \cite{afqhubb}
have indicated the presence of
anomalous flux quantization, confirming the existence of
 superconducting correlations in the ground state; In
contrast, the repulsive  Hubbard model exhibits no anomalous flux
quantization.

In addition to the existence of flux quantization, the function $E$($\Phi$)
also gives
a quantitative value of the superfluid density,
defined in one dimension by
\begin{eqnarray}
D_s={\partial^2\over \partial \Phi^2}\left
({\lim_{L\mapsto\infty}[LE_0(\Phi)]}\right )
\label{rhos}
\end{eqnarray}
A distinction should be made between $D_s$ (the superfluid density) and $D$
(the Drude weight) \cite{swz}:
The  superfluid density corresponds to the curvature of the
envelope of the individual many body states as a function of flux, whilst the
Drude
weight is obtained from the curvature of a single ground state many body
energy level. In general these quantities are different \cite{note1}.
 In the thermodynamic limit no particular
applied flux is preferred when calculating $D_s$ \cite{fye} and hence we
consider
the curvature of the whole $L$$E_0$($\Phi$) curve when considering the
superfluid density.
Note that the existence of superconductivity
requires both anomalous flux quantization and
a finite superfluid density (and infact $D_s$ has no real meaning in the
absence of anomalous flux quantization).

Numerically, the application of a flux through the double chain ring is
achieved by
modifying the kinetic term of the hamiltonian such that
\begin{eqnarray}
   {c}^{\dagger}_{j,\beta;s}{c}_{j+1,\beta;s} \mapsto
   {c}^{\dagger}_{j,\beta;s}{c}_{j+1,\beta;s} e^{i2\pi\Phi\over L}
\label{guage}
\end{eqnarray}
where $\Phi$ is the flux through the ring measured in units of $\Phi_0$ and $L$
is the
length of either chain. Hence the application of a flux is numerically
equivalent to
a change in the boundary conditions of the problem; $\Phi$=$0$ representing
periodic boundary conditions and $\Phi$=${1\over 2}$ representing anti-periodic
boundary conditions.

The technique we have employed is exact diagonalisation of finite systems,
specifically
$2\times 5$ and $2\times 10$  double chain  rings with intermediate electron
densities
$\big <n\big >=0.8$ and
$\big <n\big >=0.4$ corresponding to the regions of the phase diagram   where
we expect spin gapped or Luttinger liquid behaviour respectively.
 Note that the electron number is always a multiple of 4 in order to
guarantee that antiferromagnetic correlations are not frustrated when one goes
around each chain.
The modes of the system are characterized firstly  by their spin: singlet and
triplet excitations correspond to charge and spin modes respectively.
It is also useful to consider the parity of the states of the system under a
reflection in the symmetry axis
of the ladder along the
direction of the chains: Even ($R_x=1$) or odd ($R_x=-1$) excitations
correspond to bonding or anti-bonding modes respectively.
Finally, it is necessary to consider the momentum, $k_x$=$2\pi n/L$ in order to
determine the dispersion relation of each mode.
Implementation of these  quantum numbers and symmetries is straightforward
using exact diagonalisation methods and  the various excitation modes
may be   obtained by calculating the ground state energy in each symmetry
sector.

Concentrating initially on $J$/$t$$=0.5$ $\big <n\big >=0.8$ ,
we show in Figs.\ \ref{f1}a(b)
  all possible spin and charge modes of the $2\times 5$ ($2\times 10$) system,
  for all possible
momenta,
as a function of applied flux.
In  the case of the larger system,
we have omitted some of the details of the excited states to
simplify the diagram, showing the full spectrum only for
 $\Phi < 0.25$.
For both system sizes, the minimum energy function $E_0$($\Phi$)
is formed by charge (spin zero) bonding modes;
the excited modes with different quantum
numbers  move further from the ground state as the system size increases (a
result we have checked by finite size scaling techniques) and
hence will not interfere with  $E_0$($\Phi$).
The existence of minima at intervals of half a flux quantum (i.e.
anomalous flux quantization) clearly
indicates the existence of pairing.

In order to probe the behaviour of
 $E_0$($\Phi$) further, we consider the quantity
 $L \left [{E_0(\Phi)-E_0(\Phi =0)}\right ]$ as a function of $\Phi$
 for various values of $J/t$ and $\big < n\big >$ (L is the length of the
ladder).
Note that the curvature of this function in the thermodynamic
limit gives the superfluid density.
Fig.\ \ref{f2}a shows the contrasting behaviour obtained when
keeping $J/t$ fixed at $1.0$
and varying
the electron  filling, specifically
$\big <n\big >=0.4$ and
$\big <n\big >=0.8$ (both the $2\times 5$ and
$2\times 10$ results are shown).
This plot  clearly
shows the existence of  anomalous flux quantization
for a filling of
$\big <n\big >=0.8$ and its absence for
$\big <n\big >=0.4$.
The occurence of the absolute minima at different values of flux
($\Phi =0$ and $\Phi =1/2$ for
$\big <n\big >=0.8$ and
$\big <n\big >=0.4$ respectively)
can be explained by considering the non-interacting Fermi sea for the two
fillings;
a lower energy state is formed by choosing the flux (and hence
boundary conditions) to give a closed shell.
Fig.\ \ref{f2}b shows an equivalent plot but in this case keeping
$\big <n\big >$ constant at $0.8$ and varying the parameter $J/t$ from $0.5$
to $4.0$.
In this case
   anomalous flux quantization is exhibited  for $J/t$=0.5, whilst for
$J/t$=4.0
 $L \left [{E_0(\Phi)-E_0(\Phi =0)}\right ]$
 appears to  scale to a flat function, consistent with the
 existence of a phase separated region ($D_s=0$).

Except for the region believed to be phase separated,
the form of the curve
$L \left [{E_0(\Phi)-E_0(\Phi =0)}\right ]$
appears to show only relatively small finite size effects and
may be easily extrapolated to the thermodynamic limit;
hence, by considering the  curvature, an accurate value of the superfluid
density
(and the Drude weight)
can be found.
In a future publication \cite{chdp}
we analyse the specific values in more detail but
for this letter we emphasise that $D_s$ scales
to a finite value in the thermodynamic limit
in the regions which are not phase separated.

\bigskip
The second form of evidence for a superfluid density lies in the dispersion
behaviour of the
low energy spin and charge excitations  of the system.
The possible phases of a particular  model can be characterized
by the number of charge and spin modes which are gapless at zero momentum.
The one-dimensional analog of a superconductor has
one gapless charge mode, a gap to all spin excitations and dominant pairing.
The phase diagram for the
various possible phases for a chain with a Hubbard (rather than t--J)
hamiltonian
has recently been found \cite{fisher} and interestingly  on  doping away
from half filling, this spin gapped phase with one gapless charge mode is
stabilised, exactly the
situation we will  suggest for the t--J case.
In addition, both this article and a recent paper by Nagaosa \cite{nagaosa}
note that
in this gapped phase, in contrast to a Luttinger liquid phase,
the power law term in the density-density correlation function
at $2k_f$ is missing.

In Fig. \ref{f3} we show the dispersion
of all the spin and charge modes (for both
bonding and antibonding symmetry sectors),
 for the $2\times 10$ ladder with
 $\big < n\big >=0.8$, $J/t=1.0$, corresponding to the region of the phase
diagram
where we believe the superfluid to exist.
There are several obvious features of this diagram. Firstly there is a finite
gap to
spin excitations; secondly there is (at least) one vanishing charge mode
(bonding)
as $k_x$$\mapsto$0 and thirdly, there is no sign of $2k_f$  charge gapless
modes.
All of these features point towards a spin liquid with dominant superconducting
correlations  as previously  explained.
We should note the possibility of a  new charge gapless mode at finite momenta
corresponding to the fluctuations of the pair density as indicated by the dip
of the bonding charge mode in Figure \ref{f3}.
Some more detailed descriptions of these results (along with the corresponding
results
from other regions of the phase diagram) are given in a separate publication
\cite{DP}.

\bigskip

The final results we present are direct  numerical calculations of various
correlation functions obtained  using the
density matrix renormalization group approach \cite{rng}; this technique
 allows much larger systems to be considered than
is possible with exact diagonalization techniques.
We present three types of correlations: Firstly the equal-time rung-rung pair
field
correlation function $\big < \Delta_{i} \Delta_j^\dagger\big >$ where
$\Delta_j^\dagger=(c_{j,1;\uparrow}^\dagger c_{j,2;\downarrow}^\dagger -
c_{j,1;\downarrow}^\dagger c_{j,2;\uparrow}^\dagger)$,
$i$ and $j$ are rung indices and
$1$ and $2$ indicate the chain;  this correlation function then
creates a singlet pair on rung $j$ and removes a singlet pair from rung $i$, a
direct
measure of the motion of hole pairs in the spin gapped state.
The second correlation function that we consider is
the spin-spin correlation function ${\bf S}^z_i\cdot {\bf S}^z_j$ along
one of the chains.
Finally we measure the density-density correlations along one of the chains
defined
by $\big < \rho_{i,1}\rho_{j,1}\big >$
 where $\rho_{i,1}=\sum_{\sigma}c_{i,1;\sigma}^\dagger c_{i,1;\sigma}$.

In Fig. \ref{f4}  we show the results of a
direct calculation of these  correlation functions for a
$2\times 30$ ladder with open boundary conditions and
$\big <n\big >=0.8$ and $J/t=1.0$ i.e.  corresponding to the
region where we expect superfluid behaviour.
The data is shown on a log-log plot and the dashed line
corresonds to a slope of $-2$, the
dotted to a slope of $-1$.
The most immediate feature of the results is the fact
that the pairing correlations are
much longer range than the other correlation functions,
decaying with behaviour slower than
$1/L$. The density-density charge correlations appear to decay close to $1/L^2$
whilst
the spin-spin correlations are consistent with an exponential decay (as we
would expect for the spin gapped state).
These results are then further evidence for our
interpretation of the behaviour in this
region of the phase diagram:
dominant superconducting correlations with a spin singlet (gapped)
wavefunction.

\bigskip
In summary, we have presented three independent forms of evidence
for a superfluid density
in hole doped t--J ladders.
Firstly we have shown the existence of anomalous flux quantization and a
well-converged and finite
$D_s$; Secondly we have studied the low lying modes, finding a spin-gapped
state with
a gapless charge mode and no gapless $2k_f$ excitations; Finally we have
presented direct calculations of correlation functions, showing long range
pairing correlations and exponentially decaying spin correlations.

\bigskip
We gratefully acknowledge many useful discussions and comments from
M. Luchini and F. Mila.
{\it Laboratoire de Physique Quantique, Toulouse} is
{\it Unit\'e de Recherche Associ\'e au CNRS No 505}.
CAH and
DP  acknowledge support from the EEC Human Capital and Mobility
program under Grants ERBCHBICT941392 and  CHRX-CT93-0332; DJS  acknowledges
support from the National Science Foundation
under Grant No. DMR92-25027.
We also thank IDRIS (Orsay)
for allocation of CPU time on the C94 and C98 CRAY supercomputers.

%
%
\begin{figure}
\caption{
Energy as a function of flux (in units of $\Phi_0$=$hc/e$) for different system
sizes
with $J/t$$=0.5$,$\big <n\big >=0.8$.
We show all possible momenta  for various quantum numbers:
For the charge modes  the solid lines correspond to bonding and the dotted
lines to
anti-bonding, whilst for the spin modes the dashed lines correspond to bonding
and
the dot-dashed lines to anti-bonding.
 Figure 1a (1b)
corresponds to a system size of $2\times 5$ ($2\times 10$); for the larger
system
size we give only the charge bonding modes and the lowest lying spin
anti-bonding
mode in full in order  to simplify the diagram.
\label{f1}
}
\end{figure}

%
%
\begin{figure}
\caption{
 $L \left [{E_0(\Phi)-E_0(\Phi =0)}\right ]$
where L is the length of the ladder  and $E_0$($\Phi$) is the ground state
energy
with an applied flux $\Phi$.  The dashed lines correspond
to $2\times 5$, the solid lines to $2\times 10$. Fig.2a shows the results for
$\big <n\big >=0.4$ and
$\big <n\big >=0.8$ both with $J/t=1.0$, whilst fig.2b shows
the results for $J/t=0.5$ and $J/t=4.0$ both with
$\big <n\big >=0.8$.
\label{f2}
}
\end{figure}

%
%
\begin{figure}
\caption{
Spin and charge excitation modes of a $2\times 10$ ladder versus momentum
$k_x$ (in units of $\pi$); $\big <n\big > =0.8$ and $J/t=1.0$.
The quantum numbers associated with the various symbols
are shown on the plots.
\label{f3}
}
\end{figure}

%
%
\begin{figure}
\caption{
Log-log plot of various correlation functions versus $\mid i-j\mid$ (real space
separation) for a $2\times 30$ open chain with
$\big <n\big > =0.8$ and  $J/t=1.0$. The dashed line has a slope $-2$ and the
dotted
line $-1$. The correlation function are explained in the main text.
\label{f4}
}
\end{figure}


\begin{references}
\bibitem{note1}
 In  one-dimension
 (and in  coupled-chains)
 $D$ and $D_s$ are equivalent  since in the thermodynamic limit there are
only a finite number of energy level crossings \cite{swz};
a way to interpret this behaviour
is that {\it all} electrons which are superconducting are also conducting.
\end{references}
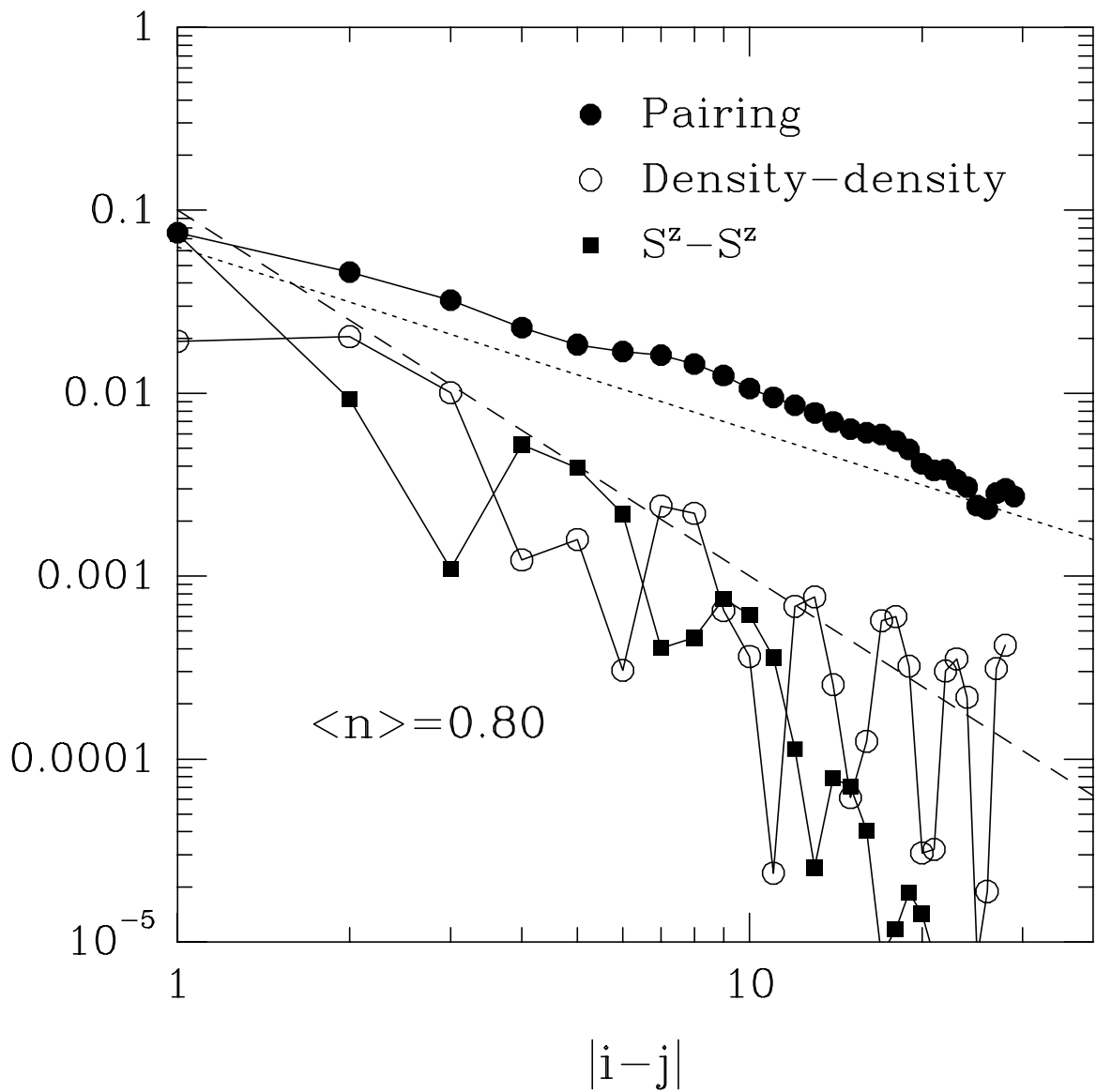
\end{document}